\documentclass[twocolumn,showpacs,prb,amsfonts,amsmath,amssymb,floatfix]{revtex4}
\usepackage{hhline}
\usepackage{mathrsfs}
\usepackage{graphicx}
\usepackage{dcolumn}
\usepackage{bm}

\input{epsf}

\newcommand{\bmk}{\bm k}
\newcommand{\ep}{\epsilon}
\newcommand{\om}{\omega}
\newcommand{\Si}{\Sigma}
\newcommand{\si}{\sigma}

\newcommand{\iomn}{i\om_n}

\newcounter{ichi}
\newcommand{\K}{\mbox{K}}
\newcommand{\eV}{\mbox{eV}}

\newcommand{\dd}{\mbox{d}}

\newcommand{\lbar}{\mbox{\underline{L}}}
\newcommand{\eg}{\mbox{e}_{g}}
\newcommand{\tg}{\mbox{t}_{2g}}

\newcommand{\aatg}[1]{^#1\mbox{A}_{2g}}
\newcommand{\ttog}[1]{^#1\mbox{T}_{1g}}

\newcommand{\eeg}[1]{^#1\mbox{E}_{g}}
\newcommand{\egg}[1]{\mbox{e}^{#1}_{g}}
\newcommand{\tgg}[1]{\mbox{t}^{#1}_{2g}}
\newcommand{\fmlda}{\mbox{\scriptsize{LDA}}}
\newcommand{\fmHF}{\mbox{\scriptsize{HF}}}

\newcommand{\Ghana}{\mathscr{G}}

\newcommand{\scr}[1]{\scriptsize{\mbox{#1}}}

\newcommand{\fr}[2]{\frac{#1}{#2}}

\newcommand{\up}{\uparrow}
\newcommand{\dw}{\downarrow}
\newcommand{\ry}{\rightarrow}

\begin{document}
\preprint{cond-mat/0110430}

\title{
Electronic structure and effects of dynamical electron correlation 
in ferromagnetic bcc-Fe, fcc-Ni 
and antiferromagnetic NiO
}

\author{Oki Miura$^{1,2}$}
\email{miura@coral.t.u-tokyo.ac.jp}
\author{Takeo Fujiwara$^{1,2,3}$}
\email{fujiwara@coral.t.u-tokyo.ac.jp}
\affiliation{$^1$Department of Applied Physics, University of Tokyo,   
Bunkyo-ku, Tokyo 113-8656, Japan} 
\affiliation{$^2$Core Research for Evolutional Science and Technology, 
Japan Science and Technology Agency (CREST-JST), Kawaguchi, 332-0012 Japan}
\affiliation{$^3$Center of Research and Development for Higher Education, 
University of Tokyo, Bunkyo-ku, Tokyo 113-0033, Japan} 


\begin{abstract}
We have constructed a 
LDA+DMFT method in the framework of the iterative perturbation theory (IPT) 
with full LDA Hamiltonian without mapping onto the effective Wannier orbitals. 
We then apply this LDA+DMFT method to ferromagnetic bcc-Fe and fcc-Ni 
as a test of transition metal, 
and to antiferromagnetic NiO as an example of transition metal oxide. 
In Fe and Ni, the width of occupied 3d bands is narrower than those in LDA 
and Ni $6~\eV$ satellite appears. 
In NiO, the resultant electronic structure is of charge-transfer insulator type 
and the band gap is $4.3~\eV$. 
These results are in good agreement with the experimental XPS. 
The configuration mixing and dynamical correlation effects play a crucial role in these results. 
\end{abstract} 
\maketitle

\section{Introduction}

Much attention has been focused on strongly correlated electron systems,  
since these systems show 
anomalous physical properties 
such as 
various spin, charge and orbital order, metal-insulator transition and so on. 
The local density approximation (LDA) based on the density functional theory (DFT) 
is hardly applicable to these fruitful physical properties in 
strongly correlated electron systems. 
LDA+U method could not treat metallic phase near metal-insulator transition 
since it has been developed in order to discuss magnetic insulators.~\cite{re:LDA+U2} 
Fluctuation of charge or/and spin densities plays an important role 
in low-energy excitations near the metal-insulator transition. 
To discuss realistic materials of strongly correlated electron systems, 
another method is needed for the ``dynamical'' electron correlation.

The dynamical mean field theory (DMFT) has been developed and applied 
to model systems of strongly correlated electrons, 
which led us to a unified picture of 
low- and high-energy excitations in anomalous metallic phase near 
metal-insulator transition.~\cite{re:Rev-DMFT} 
DMFT is based on mapping of many electron systems in bulk 
onto single impurity atom embedded in effective medium, namely the single impurity problem. 
In this mapping procedure, the on-site dynamical correlation is included and 
the inter-atomic correlation is neglected. 
The combination of DMFT with LDA, called LDA+DMFT, 
has been developed in order to discuss realistic systems.~\cite{re:Ans,re:Lic} 
LDA or DFT meet their own new stage with the GW approximation,~\cite{re:GW-Arya} 
which is based on the many body perturbation theory 
and the random phase approximation (RPA). 
The combination of DMFT with the GW approximation (GW+DMFT) was proposed 
to include both on-site and inter-atomic Coulomb interaction.~\cite{re:GW}

To solve the mapped single impurity problem within DMFT, 
one can use several computational schemes such as 
the quantum Monte Carlo method~(QMC),~\cite{re:Rozen-QMC,re:Geo-QMC,re:QMC-difficult,re:PQMC,re:CTQMC} 
the iterative perturbation theory~(IPT),~\cite{re:Kaj-IPT,re:Fuji,re:miura,re:MOIPT} 
the non-crossing approximation~(NCA)~\cite{re:NCA} 
and the exact diagonalization~(ED).~\cite{re:Geo-ED}

Hirsch-Fye QMC~\cite{re:Hirsch-Fye-QMC} is widely used as a solver for the mapped single impurity problem within DMFT. 
However, it is not applicable in the low temperature limit. 
Moreover, Hirsch-Fye QMC has serious difficulty in application of multi-orbital systems 
with spin-flip and pair-hopping terms of the exchange interactions, 
since one cannot apply the Hubbard-Stratonovich transformation 
to these systems.~\cite{re:QMC-difficult} 
To go beyond such problems of Hirsch-Fye QMC, 
several QMC methods have been proposed. 
Projective QMC~\cite{re:PQMC} has been developed to calculate in the low temperature limit 
and Continuous Time QMC~(CTQMC)~\cite{re:CTQMC} to calculate 
multi-orbital systems with spin-flip and pair-hopping terms.

In spite of development of those QMC, 
the exact calculations such as QMC and ED can only be suitable 
for simple Hamiltonians with relatively small size because of their computational costs. 
Several LDA+DMFT methods with QMC adopt the projected effective Hamiltonian 
of Wannier-like functions with a few adopted bands 
for reducing computational costs. 
With the use of the effective Hamiltonian of Wannier-like functions in LDA+DMFT, 
only a few adopted bands are discussed with fixing the hybridization mixing 
and could not describe a possible change of hybridization 
due to local Coulomb interaction.

To carry out calculations in realistic materials with multiple-orbitals, 
approximate calculation schemes of IPT or NCA could be more suitable 
because of its efficient CPU-time, 
though IPT is not applicable to cases of large Coulomb interactions and 
NCA cannot yield the Fermi liquid behavior at low energies 
and in low temperature limit.~\cite{re:NCA-error}

The IPT method was developed by Kajueter and Kotliar for non-degenerate orbital
and the results show good agreements with the ED results.~\cite{re:Kaj-IPT} 
However, it was reported that the quasiparticle peak in the IPT results 
tends to be sharper~\cite{re:Rev-DMFT} 
and that 
the IPT results in La$_{1-x}$Sr$_x$TiO$_3$ does not reprduce quasiparticle peak  at 1000K.~\cite{re:Nekrasov-LaSrTiO3}

We generalized IPT for multi-orbital bands 
and applied this method to 
doubly degenerated $\eg$ bands 
and triply degenerated $\tg$ bands 
on simple cubic and body-centered cubic lattices.~\cite{re:Fuji, re:miura} 
We also verified that 
this DMFT with IPT scheme is fairly applicable to 
arbitrary electron occupation cases with different Coulomb interaction $U$. 
The spectrum shows electron affinity and ionization levels 
with 
different electron configurations of different occupation numbers 
and the system becomes insulating state at an integer filling in sufficiently large $U$.~\cite{re:Fuji, re:miura} 
IPT for multi orbitals was also proposed by Laad, Craco and M\"uller-Hartmann and 
applied to several realistic materials.~\cite{re:MOIPT}

In this paper, we apply this DMFT with IPT scheme to realistic materials, 
where we adopt full LDA Hamiltonian without reducing its size and 
 IPT as a solver for the mapped single impurity problem. 
In the following, we notify this DMFT with IPT as ``LDA+DMFT-IPT(1)''.   
The goal of the present paper is as follows: 

(i) To generalize LDA+DMFT-IPT(1) which is applicable to various 
realistic strongly correlated materials, 
both metallic and insulating, 
multi-atom (compound), 
spin-polarized and strongly hybridized cases between s, p and d-bands. 

(ii) To apply LDA+DMFT-IPT(1) to ferromagnetic bcc-Fe and fcc-Ni 
and to antiferromagnetic NiO. 
For ferromagnetic bcc-Fe and fcc-Ni, 
we will discuss whether LDA+DMFT-IPT(1) reproduces 
the accurate width of occupied 3d bands and 
the observed satellite of spectrum at $6~\eV$ below the Fermi energy in Ni (``$6~\eV$ satellite''). 
Application of LDA+DMFT-IPT(1) to ferromagnetic bcc-Fe and fcc-Ni is a test in transition metals 
since previous LDA+DMFT~\cite{re:Fe-Ni-UJ-paper,re:Fe-Ni-KKR+DMFT} has been applied to those metals 
and reproduced those physical properties well. 
For antiferromagnetic NiO, 
we will discuss whether LDA+DMFT-IPT(1) reproduces
the accurate band gap and the correct description of the charge-transfer insulator.
These physical properties are not well reproduced in the previous LSDA calculation.

(iii) To verify the applicability of LDA+DMFT-IPT(1) to various 
realistic strongly correlated materials. 
Actually, the physical properties mentioned in (ii) are reproduced fairly well by LDA+DMFT-IPT(1). 
We will discuss the origin of drastic changes of those physical properties 
and show the validity of LDA+DMFT-IPT(1) 
by comparing with other theoretical scheme 
including other LDA+DMFT method~\cite{re:Fe-Ni-UJ-paper,re:Fe-Ni-KKR+DMFT,re:Kunes-NiO-DMFT} and experiments. 
We will discuss the origin of nickel $6~\eV$ satellite, 
which has been concluded as hole-hole scattering process,~\cite{re:Ni-satellite} 
by using the spectrum of Ligand Field Theory (LFT) obtained in the procedure of LDA+DMFT-IPT(1). 
In bcc-Fe and fcc-Ni case, the narrowing of occupied 3d bands in LDA+DMFT-IPT(1) occurs 
due to the on-site dynamical electron correlation. 
In NiO case, the appearance of charge-transfer insulator and accurate band-gap in LDA+DMFT-IPT(1) 
is due to 
the change of hybridization between Ni-3d and O-2p bands caused by local Coulomb interaction of nickel 3d bands.

In Sec.~\ref{sec:FORM}, the general formulation of 
LDA+DMFT-IPT(1) will be given. 
In Sec.~\ref{sec:result-Fe-Ni}, we test LDA+DMFT-IPT(1) 
in systems of ferromagnetic bcc-Fe and fcc-Ni. 
In the energy spectra, one can see nickel $6~\eV$ satellite caused by hole-hole scattering process 
and the narrowing of occupied 3d bands caused by on-site dynamical electron correlation. 
Section \ref{sec:result-NiO} will be devoted to calculation 
of the electronic structures in antiferromagnetic NiO by LDA+DMFT-IPT(1). 
Local Coulomb interaction of nickel 3d bands enhances the hybridization between Ni-3d and O-2p bands and 
system becomes the charge-transfer insulator with accurate band gap. 
In both sections, we will present the energy spectra and $\bmk$-resolved spectrum 
and LFT spectra of single isolated ion.
Section \ref{sec:CONC} is the summary. 
We discuss the applicability of LDA+DMFT-IPT(1) to various 
realistic strongly correlated materials by comparing with other LDA+DMFT method and experiments.

\section{Formulation}\label{sec:FORM}
\subsection{Hamiltonian and Coulomb interactions}
We proceed with the Hubbard-type Hamiltonian as follows :
\begin{eqnarray}
\! \! \! H &=& H_{\fmlda}^{dc} + H_{int}~,
\label{eqn:ham}\\
\! \! \!  H_{\fmlda}^{dc} \! \! 
&=& \! \! \! \! \! \! \! \! \! \! \! \sum_{j(m\si)j'(m')} \! \! \! \! \! \! \! 
(h^{jj'(\fmlda)}_{mm'\si}
  -\Delta h^{\si}_{dc}\delta_{jj'}\delta_{mm'}) c_{jm\si}^{\dag}c_{j'm'\si}~,
\label{eqn:HLDAdc}\\
\! \! \!  H_{int} &=& \! \! \fr{1}{2} \! \! \sum_{j(m\si)}\! \! 
U_{m_1m_2:m_3m_4}c_{jm_1\si_1}^{\dag}c_{jm_2\si_2}^{\dag}c_{jm_4\si_4}c_{jm_3\si_3}~. 
\nonumber \\
\label{eqn:int}
\end{eqnarray}
The Hamiltonian $H$ is divided into two parts:
the unperturbed part ($H^{dc}_{\fmlda}$) and the interaction part ($H_{int}$). 
$h^{jj'(\fmlda)}_{mm'\si}$ in Eq.~(\ref{eqn:HLDAdc}) is the full LDA Hamiltonian 
constructed by the Tight-Binding LMTO (TB-LMTO) method,~\cite{re:LMTO} 
without projecting onto any kind of effective local orbitals. 
The use of original full LMTO Hamiltonian in LDA+DMFT-IPT(1) enables us to describe 
the change of hybridization caused by local Coulomb interaction. 
$\Delta h^{\si}_{dc}$ in Eq.~(\ref{eqn:HLDAdc}) is the double-counting term 
included in LDA Hamiltonian as averaged Coulomb and exchange interactions, 
because we need to subtract $\Delta h^{\si}_{dc}$ from $h^{jj'(\fmlda)}_{mm'\si}$  
in order to construct the unperturbed part 
of the Hamiltonian $H^{dc}_{\fmlda}$. 
Note that an index $j$ in the summation of Eqs.~(\ref{eqn:HLDAdc}) and (\ref{eqn:int}) 
runs over atomic sites, $\{m\}$ orbital indices and $\{\si\}$ spins.

$U_{m_1m_2:m_3m_4}$ in Eq.~(\ref{eqn:int}) is 
the on-site electron-electron interaction matrix 
and presented by using the Slater integrals $F^k$ as 
\begin{eqnarray}
U_{m_1m_2:m_3m_4} &=& \sum_{0\leq k\leq 2l}a_k(m_1m_2:m_3m_4)F^k ,
 \label{eqn:U}\\
a_k(m_1m_2:m_3m_4) &=& \fr{4\pi}{2k+1} \sum_{q=-k}^{k} \langle lm_1|Y_{kq}|lm_2\rangle \nonumber\\
&\times & \langle lm_3|Y_{kq}^{*}|lm_4\rangle,
 \label{eqn:ak}
\end{eqnarray}
where $\{|lm\rangle\}$ are the basis set of complex spherical harmonics. 
Note that Eqs.~(\ref{eqn:U}) and (\ref{eqn:ak}) are the general expression 
and invariant under any rotational operation.~\cite{re:LDA+U2} 
For d-orbitals we only need $F^0, F^2$ and $F^4$. 
The averaged Coulomb and exchange parameters $U$ and $J$ for 3d-orbitals are defined 
in terms of the Slater integrals $F^0, F^2$ and $F^4$. 
Here we regard these equations as a definition of the Slater integrals $F^0, F^2$ and $F^4$ 
in terms of $U$ and $J$: 
\begin{eqnarray}
      U &=& F^0,  \label{eqn:F0}\\
      J &=& \frac{1}{14}(F^2+F^4),  \label{eqn:F2pF4} 
\end{eqnarray}
with a ratio of $F^2$ and $F^4$ for 3d orbitals as~\cite{re:F2F4-ratio1} 
\begin{eqnarray}
F^4/F^2 &\sim& 0.625.  \label{eqn:F2F4}
\end{eqnarray}
Therefore, once we obtain experimentally or assume the values $U$ and $J$ 
of 3d transition metal elements, 
the values of the Slater integrals $F^0$, $F^2$ and $F^4$ 
can be evaluated  by using Eqs.~(\ref{eqn:F0}), (\ref{eqn:F2pF4}) and (\ref{eqn:F2F4}) 
and then each matrix element of the on-site electron-electron interaction 
$U_{m_1m_2:m_3m_4}$ can be evaluated by using Eq.~(\ref{eqn:U}) and (\ref{eqn:ak}).

As discussed above, 
we treat the interaction Hamiltonian Eq.~(\ref{eqn:int}) for full 3d-orbitals on transition metals 
within the framework of DMFT. 
Thus, the present Hamiltonian in Eq.~(\ref{eqn:int}) does work for the cases with 
partially filled $\mbox{e}_{g}$ and $\mbox{t}_{2g}$-orbitals of 
ferromagnetic bcc-Fe, fcc-Ni and antiferromagnetic NiO. 
On the other hand, the case of triply degenerated $\tg$-orbitals and 
doubly degenerated Hubbard model were treated within the framework of DMFT 
in Refs.~\onlinecite{re:MOIPT} and \onlinecite{re:2band-Hubbard}, respectively.
%

\subsection{Dynamical mean field theory}

The matrices of the lattice Green's function and the local Green's function 
are presented on the basis of the non-orthogonal local base set as
\begin{eqnarray}
&& [G(\bmk, \iomn )]^{-1}
\nonumber\\
&&\ \ \ \ =[(\iomn+\mu)O(\bmk)-\{H^{dc}_{\fmlda}(\bmk)+\Si(\iomn)\}],
\label{eqn:Gkw}\\
&& G(\iomn)= \fr{1}{V}\int d\bmk~G(\bmk,\iomn) ,
\label{eqn:Gw}
\end{eqnarray}
where $H^{dc}_{\fmlda}(\bmk)$, $O(\bmk)$ and $\mu$ are 
the Hamiltonian matrix of Eq.~(\ref{eqn:HLDAdc}), 
the overlap matrix, both in the $\bmk$-space, and the chemical potential, respectively. 
Here we neglect the $\bmk$-dependence of the self-energy $\Si$ 
within the framework of DMFT. 
The chemical potential $\mu$
is determined to satisfy Luttinger's theorem.~\cite{re:Mul} 
All the matrices in Eqs.~(\ref{eqn:Gkw}) and (\ref{eqn:Gw}) have suffices 
$\{jm\sigma, j^\prime m^\prime \sigma^\prime\}$, where $j$, $m$ and $\sigma$ stand for 
atomic sites in a unit cell, orbitals and spins, respectively.~

We assume that 
the Green's function in the effective medium $G^0(\iomn)$ may be  
expressed as the self consistent equation of DMFT:
\begin{eqnarray}
G^0(\iomn)^{-1}=G(\iomn)^{-1}+\Si(\iomn)+\tilde{\mu}-\mu,~
\label{eqn:Gw-local}
\end{eqnarray}
where $\tilde{\mu}$ is the chemical potential in the effective medium.~

\subsection{Iterative Perturbation Theory}\label{sec:IPT}
In this paper, IPT method is adopted as a solver 
for the mapped single impurity problem, generalized by Fujiwara {\it et al.} for multi-orbital bands 
on arbitrary electron occupation with different Coulomb $U$.~\cite{re:Fuji, re:miura} 
In the IPT scheme, the self-energy is determined with the interpolation scheme  
between the high frequency limit and the strong interaction limit (the atomic limit) 
by using the second order self-energy.

The second order self-energy is calculated as 
\begin{eqnarray}
\!\!\!\!\!\!\!\!\Sigma^{(2)}_{m\si}(\tau) \!\!\!\! &=& \!\!\!\! - \!\!\!\!\!\!\!\!\!\!\sum_{m_1m_2m_3\si '}\!\!\!\!\!\!\!(U_{m_1m_2:mm_3}-U_{mm_2:m_3m_1}\delta_{\si\si '})\nonumber\\
&\times& \!\!\!\! U_{m_1m_3:mm_2}G^0_{m_1\si}(\tau)G^0_{m_3\si '}(\tau)G^0_{m_2\si '}(-\tau)
\label{eqn:2nd-pertu-Sig}\\
\!\!\!\!\!\!\!\!\Sigma^{(2)}(i\omega_{n}) \!\!\!\! &=& \!\!\!\!\! \int_{0}^{\beta} \!\!\!\! d\tau 
e^{i\omega_{n}\tau}\Sigma^{(2)}(\tau) \ . \label{eqn:2nd-pertu-Sig-fourier}  
\end{eqnarray}
By using the second order self-energy $\Sigma^{(2)}(\iomn)$,~
the self-energy is then expressed in a matrix form as
\begin{eqnarray}
\! \! \Sigma(i\omega_n) \! = \! \Sigma^{\fmHF} \! \! \! 
+\! A\Sigma^{(2)}(i\omega_n)[1\! -\! B(i\omega_n)\Sigma^{(2)}(i\omega_n)]^{-1},
\label{eqn:sig-IPT}   
\end{eqnarray}
where $\Sigma^{\fmHF}$ is the Hartree-Fock energy.  
The matrices $A$ and $B(i\omega_n)$ are determined 
by requiring the self-energy $\Si (\iomn)$ to be exact 
in the high-frequency limit~$(\iomn \to \infty)$,~
and in the atomic limit~$(U\to \infty)$~\cite{re:Fuji} as
\begin{eqnarray}
\!\!\!\!\!\! A \!\! &=& \!\! \lim_{i\om_n\to\infty}\{\Sigma(i\omega_n)-\Sigma^{HF}\}\Sigma^{(2)}(i\omega_n)^{-1},
\label{eqn:Sig-IPT-A}\\ 
\!\!\!\!\!\! B(i\om_n) \!\! &=& \!\! \Sigma^{at(2)}(i\omega_n)^{-1}-\{\Sigma^{at}(i\omega_n)-\Sigma^{HF}\}^{-1}A ,
\label{eqn:Sig-IPT-B} 
\end{eqnarray}
where $\Sigma^{at}(i\omega_n)$ and $\Sigma^{at(2)}(i\omega_n)$ in Eq.~(\ref{eqn:Sig-IPT-B}) are the self-energy 
and the second order self-energy in the atomic limit.~
These values are calculated exactly 
within the framework of the ligand field theory (LFT), 
which is discussed in Section~\ref{sec:LF}.~

The self-energy in Eq.~(\ref{eqn:sig-IPT}) also includes 
spin-flip and pair-hopping terms of the exchange interactions 
since the second order self-energy in Eq.~(\ref{eqn:2nd-pertu-Sig}) includes those terms. 
In addition, the IPT can be achieved within efficient CPU time 
since the self-energy in Eq.~(\ref{eqn:sig-IPT}) is obtained by only matrix calculation 
and hence the IPT is applicable to realistic materials with larger size by using full-LDA Hamiltonian. 
Form the discussion in this subsection, present IPT is much appropriate impurity solver and 
present IPT appropriately treats on-site dynamical correlation effects 
including both hole-hole and electron-electron scattering 
as well as electron-hole scattering.

In the following Sections (Secs.~\ref{sec:result-Fe-Ni} and \ref{sec:result-NiO}), 
we assume the cubic symmetry around transition metal ions 
and, therefore, the self-energy can be assumed as 
\begin{eqnarray}
   \Sigma_{jm\sigma j'm'\sigma '}(i\omega_n)=
   \Sigma_{jm\sigma}(i\omega_n)\delta_{mm'}\delta_{jj'}\delta_{\sigma\sigma '}.
\label{eqn:sig-scalar}   
\end{eqnarray}
For a cubic symmetry, the local Green's function becomes diagonal 
after the $\bmk$-integration.

\subsection{Calculation in the atomic limit based on the ligand field theory}\label{sec:LF}
The Hamiltonian mapped onto an isolated atom is defined as 
%
\begin{eqnarray}
 H_{atom}      &=& H^{dc}_{atom}+H_{int} \ ,
\label{eqn:HAM-atom}\\
 H^{dc}_{atom} &=& \sum_{(m\si)}\ep_{mm'\si}c_{m\si}^{\dag}c_{m'\si} \ ,
\label{eqn:HAM-atom-dc}\\
 \ep_{mm'\si}&=&h^{(\fmlda)}_{mm'\si}-\Delta h^{\si}_{dc}\delta_{mm'} \ .
\label{eqn:ep-atom}
\end{eqnarray}
The first term $H^{dc}_{atom}$ in Eq.~(\ref{eqn:HAM-atom}) is an unperturbed 
one-electron part. 
The Coulomb interaction $H_{int}$ is the same as Eq.~(\ref{eqn:int}). 
$h^{(\fmlda)}_{mm'\si}$ is the on-site element of LDA Hamiltonian and 
therefore $h^{(\fmlda)}_{mm'\si}$ includes the information of bulk matrix 
with electron transfer, crystal field and orbital hybridization. 
$\Delta h^{\si}_{dc}$ is the double counting term same as in Eq.~(\ref{eqn:HLDAdc}). 
Then one electron energy $\ep_{mm'\si}$ in Eq.~(\ref{eqn:ep-atom}) includes the crystal field splitting. 
Once we obtain $\ep_{mm'\si}$ and the Slater integrals, 
many-electron eigenstates of Eq.~(\ref{eqn:HAM-atom}) should  be obtained 
within the framework of the ligand field theory (LFT).~\cite{re:LFT} 
Full configuration interaction (CI) calculation 
is carried out by using the basis of all the multiplet of d electrons, 
the number of which is $2^{10}=1024$. 
Using Full-CI calculation yields an exact solution for local self-energy of Eq.~(\ref{eqn:sig-IPT}) in the atomic limit. 

The Green's function, the self-energy and the second-order self-energy 
in the atomic limit are then defined as
\begin{eqnarray}
G^{at}_{mm'\si}(i\omega_{n})
&=& \fr{1}{Z}\sum_{\eta \nu } 
\fr{\langle \eta|c_{m\si}|\nu\rangle \langle \nu|c^{\dag}_{m'\si}|\eta\rangle}
{i\om_n +(E_{\eta}-\mu N_{\eta})-(E_{\nu}-\mu N_{\nu})}
\nonumber\\
&\:\:& \:\: \:\: \:\: \:\:   \times \{e^{-\beta( E_{\eta}-\mu N_{\eta})}+e^{-\beta(E_{\nu}-\mu N_{\nu})}\}, 
\label{eqn:G-atom}\\
\Si^{at}_{mm'\si}(i\om)
&=& i\om-(\ep_{mm'\si}-\mu)-[G^{at}(i\om)]^{-1}_{mm'\si}, 
\label{eqn:Sig-atom}\\
\Sigma^{at(2)}_{m\si}(i\omega) 
&=& \sum_{m_1m_2m_3\si '}(U_{mm_2:m_1m_3}-U_{mm_2:m_3m_1}\delta_{\si\si '})\nonumber\\
\times && \! \! \! \! \! \! U_{m_1m_3:mm_2}\nonumber\\
\times && \! \! \! \! \! \! \fr{ f_{m_2\si '}(1-f_{m_1\si}-f_{m_3\si '})+f_{m_1\si}f_{m_3\si '}}{i\om - \ep_{m_1\si} - \ep_{m_3\si '} + \ep_{m_2\si '} +{\tilde{\mu}}} \ ,
\label{eqn:sig2-atom}
\end{eqnarray}
where $|\eta\rangle$, $E_{\eta}$, $N_{\eta}$, $f_{m\si}$ are 
the many-electron eigenstates in the Hilbert space of $2^{10}$ multiplets,  
the energy eigenvalue, the total electron number of the eigenstate $|\eta\rangle$ 
and the Fermi distribution function obtained from full-CI calculation 
within the framework of LFT.

One can assign, 
with a help of the atomic spectrum in LFT, 
the origin of peaks in LDA+DMFT spectra, 
e.g. the initial and final multiplets of the corresponding transitions. 
Generally, peak structures in LFT spectra would suggest 
the existence of strong scattering channels in LDA+DMFT spectra 
at around these particular energies. 

From discussion in Sections \ref{sec:IPT} and \ref{sec:LF}, 
the IPT is a appropriate approximation method and 
is worth being extended to the application of realistic materials in following four reasons: 
(1) Efficient computational cost
(2) Applicability of realistic materials with larger size and more complicated 
hybridization by using full-LDA Hamiltonian [This is followed by (1)]
(3) Care of spin-flip and pair-hopping terms of the exchange interactions in multi-orbital systems
(4) Direct comparison of LFT spectra within the IPT method with the LDA+DMFT spectra 
and assignment of the origin of peaks in LDA+DMFT spectra

\subsection{Energy spectrum and $\bmk$-resolved spectrum}\label{sec:Eq-Gw-Gkw}
The energy spectrum and the $\bmk$-resolved spectrum 
may be more physically understandable if we present the Green's function 
in the orthogonal base,  
subject to the sum rule of the energy integration of the local Green's function.~
The lattice Green's function and the local Green's function 
in the orthogonal base are defined as
\begin{eqnarray}
\!\!\!\!\!\!\!\!\![\Ghana(\bmk, \om )]^{-1}\!\!\!\! &=& \!\!\! O^{-\frac{1}{2}}(\bmk)G^{-1}(\bmk, \om )O^{-\frac{1}{2}}(\bmk)
\label{eqn:Gkw-ortho-1}\\
&=&\!\!\! [(\om+\mu)
\nonumber\\
&-&\!\!\! O^{-\frac{1}{2}}(\bmk)
\{H^{dc}_{\fmlda}(\bmk)+\Si(\om)\}O^{-\frac{1}{2}}(\bmk)],
\label{eqn:Gkw-ortho-2}\\
\Ghana(\om)&=& \fr{1}{V}\int d\bmk~\Ghana(\bmk,\om).~
\label{eqn:Gw-ortho-1}
\end{eqnarray}
The energy spectrum will be shown with the imaginary part of $\Ghana(\om)$. 
$\bmk$-resolved spectrum will be also presented with  
an imaginary part of $\Ghana(\bmk,\om)$,  
which may be 
compared with the angle-resolved photoemission spectra~(ARPES).

\subsection{Computational Details}\label{sec:Comp-Detals}
The $8 \times 8 \times 8$ $\bmk$-points are used in the $\bmk$-integration 
within the whole Brillouin zone. 
The $\bmk$-integration is carried out 
by using a generalized tetrahedron method.~\cite{re:Fuji}~
The Pad\'{e} approximation is adopted for analytic continuation of 
the Green's function from the Matsubara frequencies $\iomn$ to the real $\om$-axis.~
We adopt $2^{11}=2048$ Matsubara frequencies.   

Here, all the calculation were carried out at $T=1000~\K$ 
to reduce the number of adopted Matsubara frequency. 
To consider temperature dependence of magnetic order transition, 
one should include entropy calculation as well as total energy calculation.~\cite{re:Biermann-Ce} 
Here, we do not include entropy calculation within LDA+DMFT scheme. 
Thus temperature $T=1000\K \sim 0.1\eV$ causes the spectrum broadening by at most $0.1~\eV$ 
and that do not  change the magnetic order.

We first carry out LDA calculation to obtain input LDA Hamiltonian 
and then carry out DMFT calculation. 
We adopt the converged DMFT results as output. 
We do not feed back the density by DMFT to the LDA calculation.

In NiO case, on-site self-energy of nickel 3d bands on each sublattice 
is only included and inter-atom self-energy between 
nickel atoms on different sublattices are neglected. 
Thus, our calculation is ``single-site'' DMFT, but not cluster-DMFT.~\cite{re:rev-cluster-DMFT}

\section{Ferromagnetic bcc-Fe and fcc-Ni}\label{sec:result-Fe-Ni}
LSDA calculation has achieved a great success on overall understanding of 
the ground state electronic structures of crystalline transition metals, 
including bcc-Fe and fcc-Ni. 
However, LDA overestimates the width of occupied 3d bands 
for both ferromagnetic bcc-Fe and fcc-Ni  
and could not produce the observed satellite of spectrum 
at $6~\eV$ below the Fermi energy in Ni (``$6~\eV$ satellite'').  
On the other hand, 
LDA+DMFT calculation~\cite{re:Fe-Ni-UJ-paper,re:Fe-Ni-KKR+DMFT} gives 
the width of occupied 3d bands for ferromagnetic bcc-Fe and fcc-Ni 
and the Ni $6~\eV$ satellite, 
in good agreement with XPS experimental results.~\cite{re:Fe-expt-XPS1,re:Ni-expt-XPS1}

The aim of the calculation for ferromagnetic bcc-Fe and fcc-Ni in this section 
is to see whether LDA+DMFT-IPT(1) scheme can reproduce a reasonable width of 
occupied 3d bands and Ni $6~\eV$ satellite. 
In fact, we will show that 
LDA+DMFT-IPT(1) presents the width of occupied 3d bands for bcc-Fe and fcc-Ni 
and the $6~\eV$ satellite in fcc-Ni to be in good agreement with  
both experimental ARPES~\cite{re:Fe-expt-XPS1,re:Ni-expt-XPS1} 
and also with previous LDA+DMFT results.~\cite{re:Fe-Ni-UJ-paper,re:Fe-Ni-KKR+DMFT}

\subsection{Hamiltonian and $U$, $J$ values}
Lattice constants, atomic sphere radii in the TB-LMTO method 
and the averaged values of Coulomb and exchange interactions, $U$ and $J$, 
for ferromagnetic bcc-Fe and fcc-Ni 
to construct the Hamiltonian are summarized 
in Table~\ref{tab:parameters_and_results}. 
The values of $U$ and $J$ are obtained  
with the constraint-LDA calculation.~\cite{re:Fe-Ni-UJ-paper} 
The basis set of the Hamiltonian consists of full valence bands 
of 4s, 4p and 3d orbitals, totally nine orbitals.

\begin{figure*}
  \begin{center}
   \resizebox{150mm}{!}{\includegraphics[width=210mm,height=297mm]{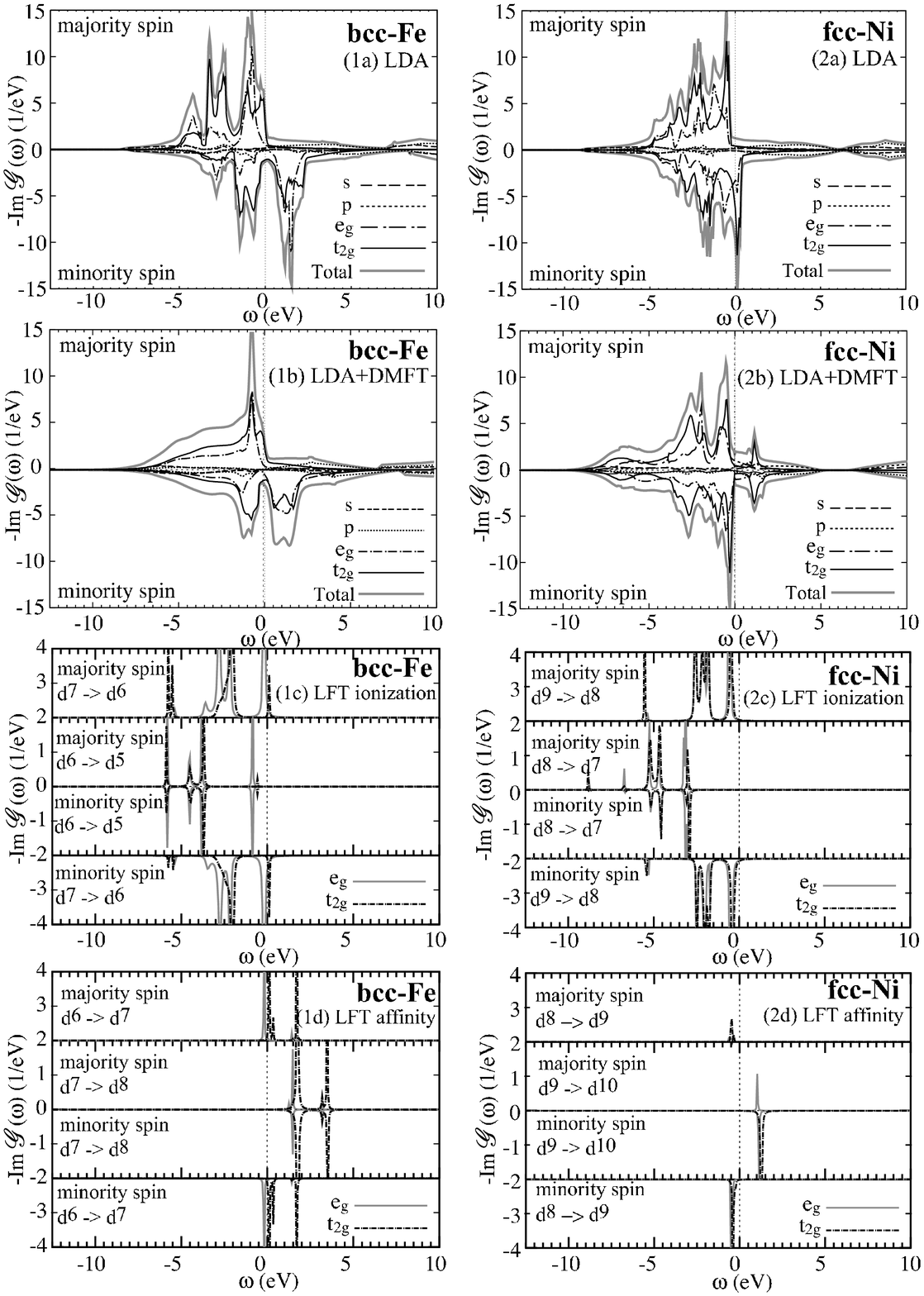}}
   \caption{
Left (1a)-(1d): Energy spectrum in ferromagnetic bcc-Fe. 
Right (2a)-(2d): Energy spectrum in ferromagnetic fcc-Ni. 
(1a)(2a): LDA. 
(1b)(2b): LDA+DMFT-IPT(1). 
(1c)(2c): Atomic ionization spectrum by the LFT. 
(1d)(2d): Atomic affinity spectrum by the LFT. 
The energy zeroth  is set at the Fermi level  ($E_{F}=0$), 
and temperature is set to be 1000~K ($T=1000~\mbox{K}$). 
}
\label{fig:dos-Fe-Ni}  
  \end{center}
\end{figure*}

\begin{table*}[]
\vspace{-5mm}
\caption{Lattice constants $a$, atomic spheres radii 
$s_0$, 
the Coulomb and exchange interactions $U$ and $J$ 
for ferromagnetic bcc-Fe, fcc-Ni and antiferromagnetic NiO. 
}
\label{tab:parameters_and_results}
\begin{center}
\begin{ruledtabular}
\begin{tabular}[t]{ccccc} 
      & lattice constant $a$ (\AA)~~ & atomic sphere radius $s_0$ (\AA) & $U$ (eV) &$J$ (eV)  \\ \hline
fcc-Ni & 3.5233   &   1.3768 &  3.0\cite{re:Fe-Ni-UJ-paper} & 0.9\cite{re:Fe-Ni-UJ-paper}  \\
bcc-Fe & 2.8708   &   1.4128 &  2.0\cite{re:Fe-Ni-UJ-paper}& 0.9\cite{re:Fe-Ni-UJ-paper}  \\
NiO    & 4.1948   & 1.2318 (Ni),  1.0699 (O),  0.8882 (ES) & 7.0\cite{re:NiO-fujimori2} & 0.9\cite{re:Anisimov-NiO-LSDA+U} \\
\end{tabular}
\end{ruledtabular}
\end{center}
\caption{The magnetic moment $\mu_{\rm spin}$ and 
the width of occupied 3d bands $W_{\rm d, occ}$ 
for ferromagnetic bcc-Fe and fcc-Ni. 
Note that $W_{\rm d, occ}$ is not shown in \onlinecite{re:Fe-Ni-KKR+DMFT}. 
}
\label{tab:Fe-Ni-past-result}
\begin{center}
\begin{ruledtabular}
\begin{tabular}[b]{ccccccc}
 \multicolumn{2}{c}{} & LSDA & GW\cite{re:yamasaki-GW-TM} & LDA+DMFT & LDA+DMFT-IPT(1) & Expt. \\
\hline
$\mu_{\rm spin}(\mu_B)$ & bcc-Fe  & 2.17 & 2.31 & 2.28\cite{re:Fe-Ni-KKR+DMFT} & 2.16 &  2.13\cite{re:Fe-Ni-expt-mag-mom} \\
                        & fcc-Ni  & 0.47 & 0.55 & 0.57\cite{re:Fe-Ni-KKR+DMFT} & 0.47 & 0.57\cite{re:Fe-Ni-expt-mag-mom} \\
\hline
$W_{\rm d, occ}(\eV)$   & bcc-Fe  & 3.7  & 3.4  &            & 3.4  & 3.3\cite{re:Fe-expt-XPS1} \\
                        & fcc-Ni  & 4.5  & 3.3  &           & 3.5  & 3.2\cite{re:Ni-expt-XPS1} \\
\end{tabular}
\end{ruledtabular}
\end{center}
\vspace{-5mm}
\caption{
The band gap $E_{\rm gap}$ (\eV) and the magnetic moment $\mu_{\rm spin}$ ($\mu_B$) 
for antiferromagnetic NiO. 
}
\label{tab:NiO-past-result}
\begin{center}
\begin{ruledtabular}
\begin{tabular}{cccccccc}
 & LSDA & LSDA+U\cite{re:Anisimov-NiO-LSDA+U} & GW\cite{re:NiO-GW-Ferdi,re:NiO-GW-Nohara} & QPscGW\cite{re:NiO-GW-Kotani} & LDA+DMFT\cite{re:Kunes-NiO-DMFT} & LDA+DMFT-IPT(1) & Expt.\\
\hline
$E_{\rm gap}(\eV) $& 0.2  & 3.7 & 0.2 & 4.8 & 4.3 & 4.3 & 4.3\cite{re:NiO-expt} \\
$\mu_{\rm spin}(\mu_B)$ & 1.00 & 1.59 & 1.00 & 1.72 & 1.85 & 1.00 & 1.64\cite{re:NiO-expt-mag-mom} \\
\end{tabular}
\end{ruledtabular}
\end{center}
\end{table*}

\begin{figure*}
  \begin{center}
   \resizebox{145mm}{!}{\includegraphics[width=297mm,height=210mm]{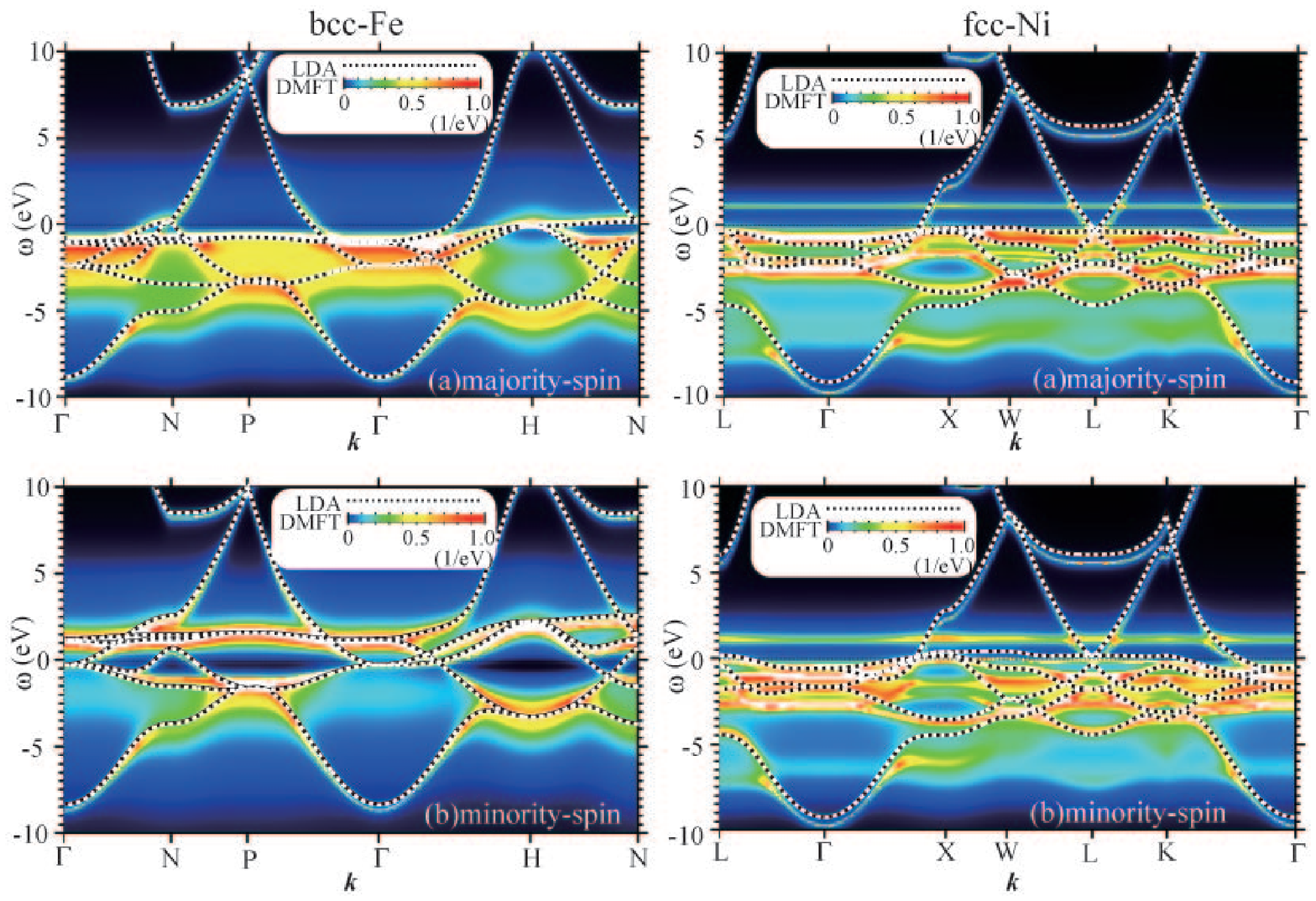}}
   \caption{The $\bmk$-resolved spectrum $-\mbox{Im} \Ghana(\bmk,\om)$ by 
LDA+DMFT-IPT(1) (shaded regions) with the energy bands (dashed lines) by LDA. 
Left:~Ferromagnetic bcc-Fe: (a)~Majority spin, (b)~Minority spin. 
The high-symmetry $\bmk$-points are $\Gamma(0,0,0)$, N$(\frac{1}{2},\frac{1}{2},0)$, 
P$(\frac{1}{2},\frac{1}{2},\frac{1}{2})$, H$(0,0,1)$. 
Right: Ferromagnetic fcc-Ni: (a)~Majority spin, (b)~Minority spin. 
The high-symmetry $\bmk$-points are L$(\frac{1}{2},\frac{1}{2},\frac{1}{2})$, 
$\Gamma(0,0,0)$, X$(0,1,0)$, W$(\frac{1}{2},1,0)$, K$(0,\frac{3}{4},\frac{3}{4})$.  
The energy zeroth is set at the Fermi energy $E_{F}=0$, 
and temperature is set to be 1000~K ($T=1000~\mbox{K}$). 
}
\label{fig:Gkw-Fe-Ni}  
  \end{center}
\end{figure*}

\subsection{Energy Spectrum}
Figure~\ref{fig:dos-Fe-Ni} shows the energy spectrum $-\mbox{Im} \Ghana(\om)$ 
obtained by LDA and LDA+DMFT-IPT(1) with the atomic spectrum by LFT. 
The occupied 3d-band becomes narrower in LDA+DMFT-IPT(1) than in LDA, 
narrower by $0.3~\eV$ and $1.0~\eV$ for bcc-Fe and fcc-Ni, respectively. 
Particularly, the spectra near the Fermi energy, 
in the region $-1.5~\eV<\om<0.0~\eV$ for both bcc-Fe and fcc-Ni in LDA+DMFT-IPT(1), 
becomes narrower than that in the low energy region. 
These results are attributed to strong renormalization of the quasiparticle 
caused by the on-site dynamical electron correlation within the framework of DMFT. 
The precise width of occupied 3d bands is discussed 
in Sec.~\ref{sec:Gkw-Fe-Ni-result}.

In addition, the satellite appears in LDA+DMFT-IPT(1) spectrum of fcc-Ni 
at $6~\eV$ below the Fermi energy, 
in good agreement with experimental XPS results.~\cite{re:Ni-expt-XPS1,re:Ni-expt-spinXPS}
Moreover, this satellite structure is strongly spin-dependent, 
much enhanced in the spectrum of the majority spin. 
These effects mainly come from the multiplet scattering of $\dd^8 \ry \dd^7$, 
which is assigned by spin-dependent peak of LFT spectrum of 
$\dd^8 \ry \dd^7$ at around $-5~\eV$ in Fig.~\ref{fig:dos-Fe-Ni}-(2c). 
The satellite is due to hole-hole scattering process,~\cite{re:Ni-satellite} 
and can appear only when one treats whole on-site dynamical correlation effects. 
It should be noted that $6~\eV$ satellite is not observed 
in GW approximation (GWA)~\cite{re:yamasaki-GW-TM} since 
GWA includes the dynamical correlation only within RPA (electron-hole excitations) 
and thus not the electron-electron and hole-hole scattering processes.

Let us focus on the atomic spectrum obtained by LFT in Fig.~\ref{fig:dos-Fe-Ni}.
The atomic spectra are separately shown 
in Figs.~\ref{fig:dos-Fe-Ni}-(1c) (2c) for ionization spectra 
and Figs.~\ref{fig:dos-Fe-Ni}-(1d) (2d) for affinity spectra. 
The initial state is a mixture of $\dd^6$, $\dd^7$ in Fe 
and $\dd^8$, $\dd^9$ in Ni, 
since the total occupation numbers are 
non-integers, $n^{\scr{Fe}}_{\rm tot}=6.829$ and $n^{\scr{Ni}}_{\rm tot}=8.799$ 
from the results of bulk systems. 
The excited state is then a mixture of $\dd^5$, $\dd^6$, $\dd^7$, $\dd^8$ in Fe and 
$\dd^7$, $\dd^8$, $\dd^9$, $\dd^{10}$ in Ni. 
Small components appear in ionization spectra above $E_F$ 
and in affinity spectra below $E_F$. 
Moreover, characteristic atomic spectra at around $-0.5~\eV$ 
in Fe and Ni originate from both ionization and affinity process; 
$\dd^6 \ry \dd^7$ and $\dd^7 \ry \dd^6$ in Fe in Figs.~\ref{fig:dos-Fe-Ni}-(1c)(1d) and 
$\dd^8 \ry \dd^9$ and $\dd^9 \ry \dd^8$ in Ni in Figs.~\ref{fig:dos-Fe-Ni}-(2c)(2d). 
These effects comes from the fact that higher energy occupied states are 
not fully occupied.

Characteristic structure in both atomic ionization and affinity spectra 
originate from various multiplet scattering; 
$\dd^7 \ry \dd^6$ and $\dd^6 \ry \dd^5$ 
at around $-4.0~\eV$ and $-6.0~\eV$ 
and $\dd^7 \ry \dd^8$ and $\dd^6 \ry \dd^7$ at around $1.5~\eV$ 
in Fe of Figs.~\ref{fig:dos-Fe-Ni}-(1c)(1d) and 
$\dd^9 \ry \dd^8$ and $\dd^8 \ry \dd^7$ at around  $-5.5~\eV$ 
in Ni of Figs.~\ref{fig:dos-Fe-Ni}-(2c)(2d). 
This is due to the multiplet scattering caused by the exchange interaction $J$. 
The primary splitting of multiplets comes from the Coulomb interaction $U$, 
for example the transition spectrum  of $\dd^8 \ry \dd^7$ should locate in the 
lower energy region than that of $\dd^9 \ry \dd^8$ of Ni spectra. 
Then additional multiplet splitting is caused by $J$. 
In LFT spectra in $-6.0~\eV < \om < 1.5~\eV $, 
the energetic order of the spectral positions does not follow 
the above first simple rule and this fact implies that 
the scattering process by $J$ changes the multiplet spectra drastically. 
We conclude that the structures in LDA+DMFT-IPT(1) spectra 
is broadened and smoothed due to 
the multiplet scattering of $\dd^7 \ry \dd^6$ and $\dd^6 \ry \dd^5$ 
at $-6.0~\eV < \om < -2.0~\eV$ and
that of $\dd^7 \ry \dd^8$ and $\dd^6 \ry \dd^7$ at around $1.5~\eV$ in bcc-Fe and 
that of $\dd^9 \ry \dd^8$ and $\dd^8 \ry \dd^7$ 
at around  $-6.0~\eV < \om <-2.0~\eV$ in fcc-Ni.

\subsection{$\bmk$-resolved spectrum and magnetic moment}\label{sec:Gkw-Fe-Ni-result}
Figure \ref{fig:Gkw-Fe-Ni} shows the $\bmk$-resolved spectrum 
$- \mbox{Im}\,\Ghana(\bmk,\om)$ 
by LDA+DMFT-IPT(1) with the LDA energy bands. 
LDA results overestimate the width of occupied 3d valence bands, 
which is defined to be the energy difference 
between the Fermi energy and the energy eigenvalue at P-point (Fe) or L-point (Ni),
in comparison with experimental results. 
The width of occupied 3d valence bands 
and the magnetic moment in LDA+DMFT-IPT(1) are shown in Table~\ref{tab:parameters_and_results}
in comparison with those by LDA and 
the experiments.~\cite{re:Fe-Ni-expt-mag-mom,re:Fe-expt-XPS1,re:Ni-expt-XPS1} 
The valence band width of LDA+DMFT-IPT(1) 
is in reasonable agreement with experiments.

One can observe flat branches in bcc-Fe  at around $-1.5~\eV$ of majority spin and 
that at around $-2~\eV$ of minority spin. 
These flat bands  correspond to the local multiplet excitations of $\dd^7 \ry \dd^6$. 
One can also observe flat branches in fcc-Ni at $-6~\eV$ 
of both majority and minority spins. 
These flat bands at $-6~\eV$ are due to the local multiplet excitations 
of $\dd^8 \ry \dd^7$, where 
the intensity of the $\bmk$-resolved spectrum of majority spin is 
larger than that of minority spin. 
This intensity difference causes the strong spin dependence 
of $-6~\eV$ satellite in the energy spectrum in Fig.~\ref{fig:dos-Fe-Ni}-(2b).

The $\bmk$-resolved spectrum in bcc-Fe is more diffusive than that in fcc-Ni at around $-3.0~\eV \sim 0.0~\eV$ 
in spite of smaller $U$ for bcc-Fe than fcc-Ni. 
This comes from that the scattering process by $J$ compared with $U$ is stronger in bcc-Fe than in fcc-Ni 
since $J/U$ in bcc-Fe is larger than that in fcc-Ni.

The magnetic moment $\mu_{\rm spin}$ of both bcc-Fe and fcc-Ni is almost 
the same as LDA result. 
$\mu_{\rm spin}$ of bcc-Fe in LDA+DMFT-IPT(1) result is also in good agreement with experiment, 
while that of fcc-Ni is slightly smaller than experimental result. 
This is an artifact due to insufficient number of $\bmk$-points to 
integrate the lattice Green's function $G(\bmk, \om)$ 
to obtain the local Green's function $G(\om)$ by using generalized tetrahedron method. 
The adopted value of the total number of the $\bmk$-points 
in LDA+DMFT-IPT(1) is much smaller than that in TB-LMTO method and 
the discrepancy of $\mu_{\rm spin}$ for fcc-Ni may be improved 
by increasing the total number of the $\bmk$-points in LDA+DMFT-IPT(1).

\subsection{Comparison with previous LDA+DMFT results}\label{sec:Compare_LDA+DMFT_for_FeNi}
Here, we compare the results for ferromagnetic bcc-Fe and fcc-Ni obtained by
LDA+DMFT-IPT(1) with other previous LDA+DMFT~\cite{re:Fe-Ni-UJ-paper, re:Fe-Ni-KKR+DMFT}. 
Lichtenstein {\it et al.}~\cite{re:Fe-Ni-UJ-paper} has used the LDA+DMFT with QMC as an impurity solver. 
Min\'{a}r {\it et al.}~\cite{re:Fe-Ni-KKR+DMFT} has used the KKR+DMFT with 
perturbative SPTF (spin-polarized T-matrix+FLEX) as an impurity solver. 
The energy spectra in LDA+DMFT-IPT(1) show a good agreement with 
those LDA+DMFT calculations. 
The existence of Ni 6~eV satellite is also very similar to 
those LDA+DMFT calculations, 
presumably much better coincident with the position of the observed spectra. 
The magnetic moment of bcc-Fe 
is in good agreement with those LDA+DMFT results. 
Slightly smaller value of the magnetic moment of fcc-Ni in LDA+DMFT-IPT(1) 
than previous LDA+DMFT is 
not due to the use of perturbative IPT approach as an impurity solver 
but due to insufficient number of $\bmk$-points mentioned above. 
Thus, we can conclude that 
LDA+DMFT-IPT(1) reproduces reasonable results for ferromagnetic bcc-Fe and fcc-Ni 
and that LDA+DMFT-IPT(1) is 
applicable to realistic metallic materials in strongly correlated electron systems 
as well as other previous LDA+DMFT methods.~\cite{re:Fe-Ni-UJ-paper, re:Fe-Ni-KKR+DMFT}

\section{Antiferromagnetic NiO}\label{sec:result-NiO}
NiO is a type-II antiferromagnetic insulator with N\'eel temperature $523~\K$. 
The experimentally observed band gap is $4.3~\eV$.\cite{re:NiO-expt} 
Resonance photoemission experiments~\cite{re:NiO-kyoumei} 
show that electronic structure of NiO should be of the charge-transfer type 
and the low energy satellite mainly consists of nickel 3d bands.

Various theoretical methods have been applied to NiO. 
The band gap $E_{\rm gap}$ and the magnetic moment $\mu_{\rm spin}$ are summarized 
in Table~\ref{tab:NiO-past-result}. 
LSDA calculation shows that 
NiO is Mott-Hubbard type insulator with a small band gap of $0.2~\eV$ 
for antiferromagnetic phase,~\cite{re:Terakura-NiO-LSDA}
in which oxygen 2p bands are located at lower energy region 
than the occupied nickel 3d bands.

The LSDA+U method was applied 
and the resultant band gap is almost good agreement with experiment 
and a system becomes charge-transfer insulator.~\cite{re:Anisimov-NiO-LSDA+U} 
However, bonding states of nickel $\eg$ bands are observed at around $8~\eV$ below the Fermi energy 
and this causes less components of the occupied main peak 
and too much components of the occupied satellite peak, 
compared with experimental XPS spectrum. 
These problems mainly come from the static potential correction with orbital dependence in LSDA+U 
and this implies that one should include dynamical correlation effects.

GW approximation was applied 
and the resultant band gap still remains very small 
(0.2\eV).~\cite{re:NiO-GW-Ferdi,re:NiO-GW-Nohara} 
Quasiparticle self-consistent GW (QPscGW) approximation~\cite{re:NiO-GW-Kotani} was also applied
and a system becomes charge-transfer insulator. 
However, the resultant band gap is overestimated (4.8\eV) 
and the bonding states of nickel $\eg$ bands 
are observed at around $6~\eV$ below the Fermi energy. 
These fact implies that 
the dynamical electron correlation should play a more crucial role in NiO than treated by RPA in GW approximation.

LDA+DMFT was applied to paramagnetic NiO \cite{re:Kunes-NiO-DMFT,re:Ren-NiO-DMFT} 
and the resultant band gap is in good agreement 
with experimentally observed XPS result.~\cite{re:NiO-expt} 
and the system becomes charge-transfer insulator. 
However, the top of valence bands mainly consists of nickel 3d bands, 
while 
the cluster-model CI calculation~\cite{re:NiO-fujimori2, re:NiO-vanElp} and 
oxygen x-ray absorption of Li$_x$Ni$_{1-x}$O~\cite{re:NiO-Kuiper-expt} show that 
the top of valence bands is mainly based on oxygen 2p bands. 
This difference comes from that these LDA+DMFT calculations use 
projected effective Hamiltonian constructed by Wannier-like functions 
with fixing the hybridization mixing. 
Moreover, no calculation has been carried out for antiferromagnetic NiO by LDA+DMFT.

The target of LDA+DMFT calculation is 
to get electronic structures of antiferromagnetic NiO with 
(i) the accurate band gap, 
(ii) the correct description of the charge-transfer type insulator and 
(iii) the occupied main peak corresponding to oxygen 2p states 
and the occupied satellite peak corresponding to nickel 3d states.

\subsection{Hamiltonian, the Coulomb and exchange interactions $U$ and  $J$}
The structure of NiO is not of the dense packing and we put empty atom spheres 
in vacant region of the lattice in LMTO formalism. 
The lattice constants, atomic sphere radii of each atom 
and the averaged values of Coulomb and exchange interactions, $U$ and $J$, 
are summarized in Table~\ref{tab:parameters_and_results}. 
The total number of atom spheres in an antiferromagnetic unit cell is 
eight; two Ni, two O and four empty atoms (ES). 
We use the full LDA Hamiltonian of the LMTO formalism and 
adopted muffin-tin orbitals are 4s,~4p and 3d in Ni and 2s, 2p in O and 
1s, 1p in ES and, thus, the total number of basis is 42. 
$U$ refers to the experimental value of $U=7.0~\eV$~\cite{re:NiO-fujimori2} 
and $J$ to the constraint-LDA calculation of $J=0.9.~\eV$~\cite{re:Anisimov-NiO-LSDA+U} 
The difference of $U$-values in fcc-Ni and NiO is 
due to the difference of screening mechanism in metals and insulators.

The value of $U$ obtained by constraint LDA is 
$U=8.0~\mbox{eV}$,~\cite{re:Anisimov-NiO-LSDA+U} 
which is slightly larger than experimental value. 
Within the framework of constraint LDA, all the screening channels are switch off. 
When we evaluate suitable value of $U$ for 3d-orbitals required 
in the framework of LDA+DMFT, 
only the screening channels associated with 3d electrons should be switch off 
and those associated with 4s and 4p electrons should remain.  
In this sense, 
a suitable value of $U$ is slightly smaller than that for constraint LDA. 
Thus, the adopted value of $U=7.0~\mbox{eV}$ in the present paper 
is reasonable within the framework of LDA+DMFT. 
%

\begin{figure*}
\vspace{-0.1cm}
  \begin{center}
   \resizebox{150mm}{!}{\includegraphics[width=297mm,height=210mm]{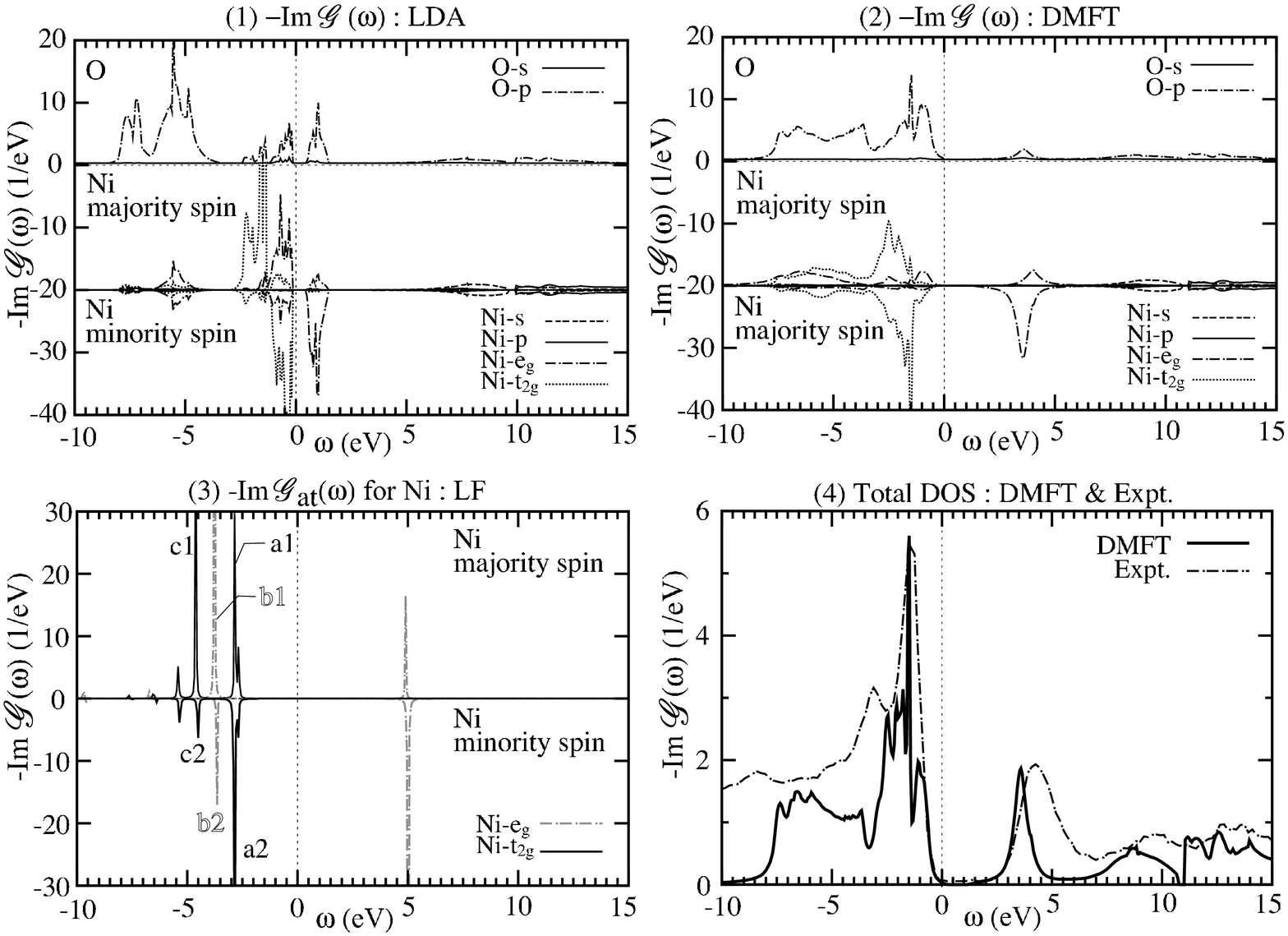}}
   \caption{Energy spectra $-\mbox{Im} \Ghana(\om)$ of antiferromagnetic NiO. 
(1) Partial $-\mbox{Im} \Ghana(\om)$ by LDA. 
(2) Partial $-\mbox{Im} \Ghana(\om)$ by LDA+DMFT-IPT(1). 
(3) Partial $-\mbox{Im} \Ghana(\om)$ in the atomic spectra of Ni by LFT. 
(4) Total $-\mbox{Im} \Ghana(\om)$ by LDA+DMFT-IPT(1) and XPS. 
The energy zeroth is set at the Fermi energy ($E_{F}=0$), and 
temperature is set to be 1000~K (T=$1000~\mbox{K}$). 
Characteristic atomic spectra for $\eg$ ($\tg$) orbitals is 
labeled by white (black) letter in (3). 
Note that the index of nickel "majority (minority) spin" denotes 
up (down) spin component on up spin Ni site  
and down (up) spin component on down spin Ni site. 
}
\label{fig:dos-NiO}  
  \end{center}

\vspace{-0.6cm}
  \begin{center}
  \resizebox{150mm}{!}{\includegraphics[width=297mm,height=210mm]{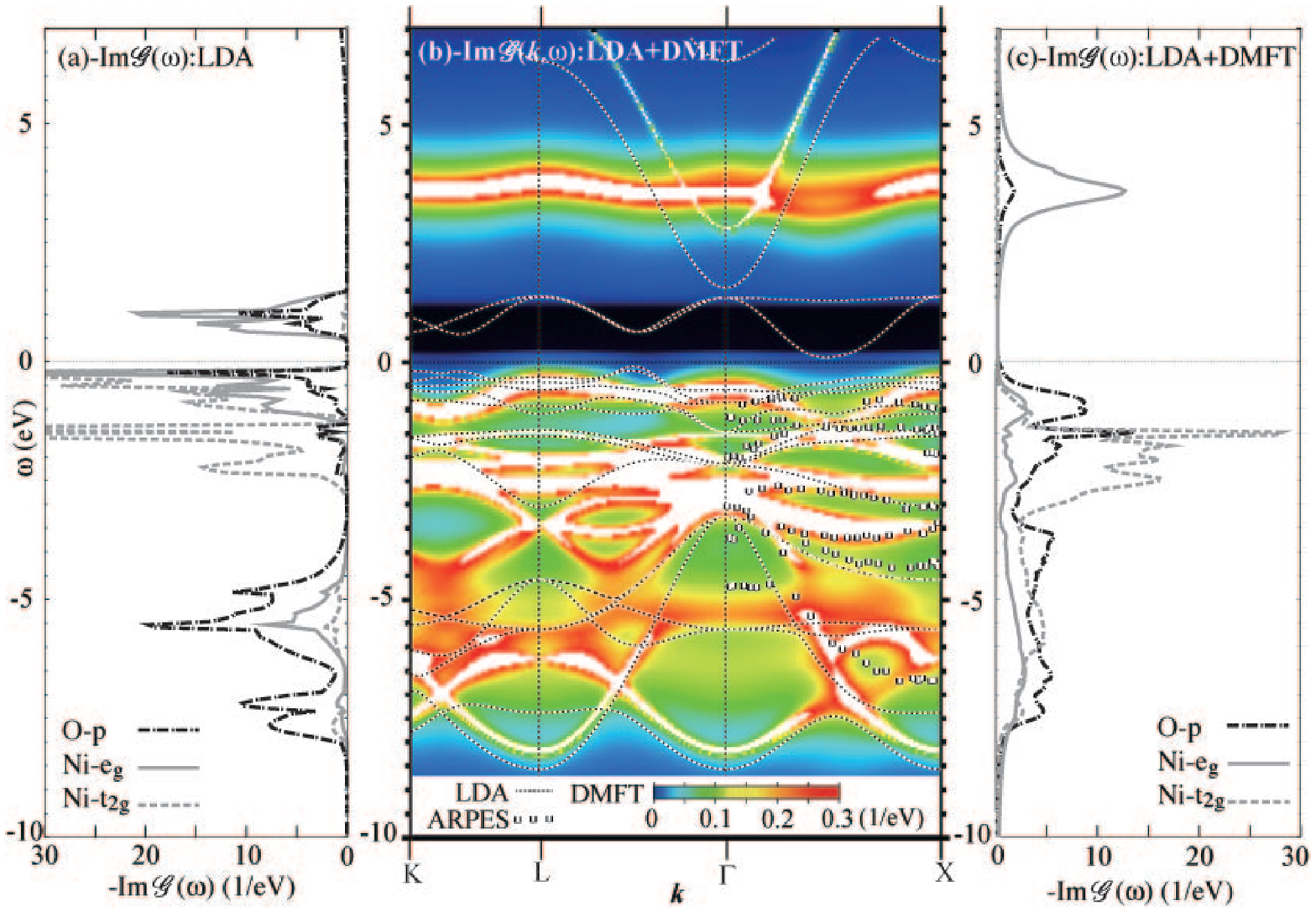}}
  \caption{(a) Partial $-\mbox{Im} \Ghana(\om)$ for antiferromagnetic NiO by LDA 
 corresponding to Fig.~\ref{fig:dos-NiO}-(1). 
 (b) $\bmk$-resolved spectrum of LDA+DMFT-IPT(1) and energy bands of LDA for antiferromagnetic NiO. 
 Shaded regions: $\bmk$-resolved spectrum $-\mbox {Im} \Ghana(\bmk,\om)$ by LDA+DMFT-IPT(1). 
 Dashed lines:~energy bands by LDA. 
 Dots:~the angle-resolved photoemission spectrum (ARPES).\cite{re:NiO-ARPES} 
 The high-symmetry $\bmk$-points are K$(0,\frac{3}{4},\frac{3}{4})$, L$(\frac{1}{2},\frac{3}{2},\frac{1}{2})$, $\Gamma(0,0,0)$, X$(0,0,1)$. 
(c) Partial $-\mbox{Im} \Ghana(\om)$ for antiferromagnetic NiO by LDA+DMFT-IPT(1) 
corresponding to Fig.~\ref{fig:dos-NiO}-(2). 
The energy zeroth is set at the Fermi energy ($E_{F}=0$), 
and temperature is set to be 1000~K ($T=1000~\mbox{K}$). 
Each figure shows the total of both majority and minority spins. 
}
\label{fig:Gkw-NiO}  
  \end{center}
\end{figure*}

\subsection{Energy Spectrum}

Figure~\ref{fig:dos-NiO} shows the energy spectrum $-\mbox{Im} \Ghana(\om)$ 
of antiferromagnetic NiO by LDA+DMFT-IPT(1) with experimental XPS spectrum.~\cite{re:NiO-expt} 
Experimental XPS spectrum mainly consists of three parts: 
a main peak at $4~\eV$, 
a main peak at $-3.5~\eV < \om < -0.5~\eV$ and 
a satellite peak at $-10.0~\eV < \om < -6.0~\eV$ 
as in Fig.~\ref{fig:dos-NiO}-(4). 
With the cluster model CI calculation,~\cite{re:NiO-fujimori2} 
these three structures are assigned to 
$\dd^9$, $\dd^8\lbar$ and $\dd^7$ final state, respectively, 
where $\lbar$ is a ligand hole created in oxygen 2p orbitals.

The energy spectrum by LDA+DMFT-IPT(1) reproduces these three structures 
fairly well,  
the positions of two main peaks are 
in good agreement with experimentally observed XPS result, 
but the position of satellite peak at $-10.0~\eV < \om < -6.0~\eV$ 
shifts slightly upward in comparison with experimental results.~\cite{re:NiO-expt} 
Appreciable component of oxygen 2p bands appears just at the top of the valence bands. 
The main peak of the occupied states at $-1.5$~eV and the satellite 
at $-6.0~{\rm eV}\sim -7.5~{\rm eV}$ originate from 
t$_{2g}$ orbitals with minority spin of Ni and 
e$_g$+t$_{2g}$ orbitals with majority spin of Ni, respectively. 
The main peak of conduction bands comes from the Ni-e$_g$ orbitals with 
minority spin. 
The spectrum of LDA+DMFT-IPT(1) in Fig.~\ref{fig:dos-NiO} shows that 
the electronic structure is of the charge-transfer insulator type, 
while the electronic structure in LDA is of Mott-Hubbard type. 
In fact, the hybridization mixing between oxygen 2p bands and nickel 3d bands 
is much enhanced in LDA+DMFT-IPT(1) in comparison with that in LDA, 
though the Coulomb matrix elements between Ni-3d and O-2p bands and 
among O-2p bands are not included. 

On-site Coulomb interaction between nickel $\eg$ bands makes 
the occupied nickel $\eg$ bands shift to the lower energy side 
and the unoccupied nickel $\eg$ bands to the higher energy side. 
Due to the shift of the occupied nickel $\eg$ bands to lower energy side, 
the hybridization between Ni-$\eg$ and Ni-$\tg$ increases 
and the occupied Ni-$\tg$ bands shift to the lower energy side. 
Since the occupied Ni-$\eg$ and $\tg$ bands shift to the lower energy side, 
the hybridization between Ni-3d and O-2p bands increases 
and whole bands are broadened in the region $-8.5~\eV<\om<-0.5~\eV$. 
These effects enhance the hybridization mixing and 
change the character of NiO 
to be of the charge-transfer insulator type. 
The energy spectrum and the electronic structure by LDA+DMFT-IPT(1) is 
in good agreement with XPS experiments.

Atomic spectrum for single nickel ion obtained by LFT 
is shown in Fig.~\ref{fig:dos-NiO}-(3). 
Initial state is $\aatg{3}(\dd^8)$ of 
the electron configuration ${\tgg{3}}_{\up}{\tgg{3}}_{\dw}{\egg{2}}_{\up}$.
The excited states of atomic ionization spectra a1 and a2 are $\ttog{4}(\dd^7)$, 
b1 and b2 are $\eeg{2}(\dd^7)$ and c1 and c2 are $\ttog{2}(\dd^7)$, respectively. 
This assignment is perfectly consistent with previous LFT calculation.~\cite{re:NiO-fujimori2}

In the atomic calculation of LFT, a single nickel ion is considered 
and inter-atomic electron transfer is not allowed. 
Therefore, present atomic calculation does not include explicitly the $\dd^8\lbar$ final state.
The initial state in the LFT spectra is fixed in $\dd^8$ 
since the total electron number of Ni ion is eight in NiO. 
Therefore, the unoccupied and occupied spectra in Fig.~\ref{fig:dos-NiO}-(3) 
originate from the transition of $\dd^8 \ry \dd^9$ and $\dd^8 \ry \dd^7$, respectively. 
Though the $\dd^8\lbar$ final state is not included in LFT, 
of Fig.~\ref{fig:dos-NiO}-(3), 
the occupied main peak in the spectrum of LDA+DMFT-IPT(1) is in good agreement with 
experimental XPS result. 
In fact, the initial states in LDA+DMFT-IPT(1) is a mixture of $\dd^8$ and $\dd^9\lbar$,  
since the electron occupation numbers of nickel 3d bands and oxygen 2p bands 
in LDA+DMFT-IPT(1) are 8.17 and 5.14, respectively. 
Therefore, we can assign, in the spectrum of LDA+DMFT-IPT(1), 
the unoccupied main peak and occupied satellite peak to be 
the $\dd^9,~\dd^7$ final state,~respectively. 
The occupied main peak is also assigned to be the $\dd^8\lbar$ final state.

The splitting between $\dd^8\lbar$ and $\dd^7$ final state configurations 
is small and, in Fig.~\ref{fig:dos-NiO}-(4), 
the occupied satellite peak in LDA+DMFT-IPT(1) appear  
at higher energy region ($ -6 \sim -7$~eV) than that in experimental XPS result 
($-8 \sim -9$~eV). 
This may be due to the fact that we do not include the $\dd^8\lbar$ final state 
in LFT calculation. 
The cluster model CI-calculation instead of LFT calculation 
would give more precise position of occupied satellite peak.~

Unoccupied $\eg$ bands of nickel majority spin locates at slightly higher energy region 
than that of minority spin and oxygen 2p bands in Fig.~\ref{fig:dos-NiO}-(1). 
It would not be the case, if we include the inter-atomic component of the self-energy 
or cluster CI-calculation.

The band gap $E_{\rm gap}$ and the magnetic moment $\mu_{\rm spin}$ are summarized 
in Table~\ref{tab:NiO-past-result}, 
The band gap in LDA+DMFT-IPT(1) is in fairly good agreement with experimental result. 
On the other hand, the calculated magnetic moment is almost the same as the 
calculated result by LDA and much smaller than the observed one. 
The magnetic moment of antiferromagnetic NiO comes from the electron occupation of 
nickel $\eg$ bands since nickel $\tg$ bands are almost fully occupied. 
The present result of smaller value of the magnetic moment in LDA+DMFT-IPT(1) than the experimental result 
comes from that unoccupied $\eg$ bands of nickel majority spin has some intensity. 
In the present calculation, we neglect the inter-atomic and inter-spin components 
of the self-energy and the discrepancy may be attributed to this approximation, 
which should be left as a future study.


\subsection{$\bmk$-resolved Spectrum}
The $\bmk$-resolved spectrum $- \mbox{Im}\,\Ghana(\bmk,\om)$ by LDA+DMFT-IPT(1) 
with both the energy bands by LDA (dashed lines) 
and the angle-resolved photoemission spectrum (dots) is shown 
in Fig.~\ref{fig:Gkw-NiO}. 
LDA band at $0.0~\eV < \om < 1.5~\eV$ is shifted 
to the LDA+DMFT-IPT(1) band at $3.0~\eV < \om < 4.0~\eV$. 
These spectra correspond to unoccupied nickel $\eg$ bands 
in Fig.~\ref{fig:dos-NiO}-(1). 
These unoccupied nickel 3d bands in LDA+DMFT-IPT(1) may be more localized 
than that in LDA.

In LDA band structure, the occupied bands mainly consists of two parts: 
(i) $-2.5~\eV < \om < 0.0~\eV$ bands structure mainly based on Ni-3d bands and 
(ii) $-8.5~\eV < \om < -3.5~\eV$ bands structure mainly based on O-2p bands.~
In LDA+DMFT-IPT(1) results, (i) and (ii) have been shifted to lower and higher energy side, 
respectively. 
This shift of these results causes the increase of hybridization 
between Ni-3d and O-2p bands. 
Especially,~strong hybridization occurs at L and $\Gamma$ points.~

We observe a broad and flat diffusive structure at $-5~\eV$ in LDA+DMFT-IPT(1) spectrum,  
which corresponds to occupied satellite peak of Ni-$\tg$ bands 
and also Ni-$\tg$ state in LFT in Figs.~\ref{fig:dos-NiO}-(2)(3). 
Therefore, we conclude that Ni-$\tg$ bands may be more localized 
due to the strong scattering by $U$.

One can see flat branches at 
$-5.5eV< \omega <-5.0~\eV$and at $ -6.5~\eV< \omega < -6.0~\eV$ 
in $\bmk$-resolved spectrum of LDA+DMFT-IPT(1). 
This corresponds to occupied satellite peak of Ni-$\eg$ bands 
in Fig.~\ref{fig:dos-NiO}-(1).~
This flat band implies 
that the occupied satellite peak of Ni-$\eg$ bands is not due to 
the band structure but to the multiplet scattering, 
particularly $\dd^8 \ry \dd^7$ assigned by the atomic calculation of LFT.

The $\bmk$-resolved spectrum at $-5.5~\eV < \om < -2.5~\eV$ of LDA+DMFT-IPT(1) is 
in good agreement  with experimental ARPES result~\cite{re:NiO-ARPES} 
along $\Gamma$ to X point, which is 
due to the inclusion of dynamical correlation within DMFT, 
which causes strong hybridization between Ni-3d and O-2p bands.

Finally, we should mention that much enhanced intensity profile can be seen in 
$- \mbox{Im}\,G(\bmk,\om)$ but $- \mbox{Im}\,\Ghana(\bmk,\om)$ becomes more diffusive 
due to overlap integrals.

\subsection{Comparison with other calculations and experiments}\label{sec:Compare_with_others_NiO}
Here, we compare the results for antiferromagnetic NiO obtained by LDA+DMFT-IPT(1) 
with other previous calculations. 
LDA+DMFT-IPT(1) reproduces 
$4.3~\eV$ band gap and the characteristics of the charge-transfer type insulator.  
Moreover, the results obtained by LDA+DMFT-IPT(1) shows 
no Ni-$\eg$ bonding states below the Fermi energy, 
which is observed in LSDA+U~\cite{re:Anisimov-NiO-LSDA+U} 
and QPscGW results.~\cite{re:NiO-GW-Kotani} 
LDA+DMFT-IPT(1) also reproduces the top of valence bands with mainly based on oxygen 2p bands,
while application of the LDA+DMFT method 
using Hamiltonian constructed by Wannier-like function 
to paramagnetic NiO~\cite{re:Kunes-NiO-DMFT,re:Ren-NiO-DMFT} shows 
that the top of valence bands is mainly based on nickel $\eg$ bands. 
Those effects obtained by LDA+DMFT-IPT(1) are in good agreement with experimental XPS results. 
Those drastic change in LDA+DMFT-IPT(1) is due to 
enhancement of hybridization between Ni-3d and O-2p bands caused by 
on-site Coulomb interaction of nickel 3d bands within DMFT scheme.

The assignment of final states of the main and satellite occupied peaks and 
the unoccupied main obtained in LDA+DMFT spectrum is consistent with 
that in the cluster-model CI calculation.~\cite{re:NiO-fujimori2} 
This comes from the use of LFT spectrum within the IPT method and hence 
the assignment of the origin of peaks in LDA+DMFT spectra by using LFT spectra 
is a great advantage of the IPT method, compared with other impurity solvers.

However we still have several problems, 
slightly higher energy positions of the occupied satellite peak and the unoccupied main peak and 
smaller value of the magnetic moment. 
From discussion in the previous two subsections, 
these are mainly 
due to neglect of the inter-atomic and inter-spin components 
of the self-energy within DMFT scheme. 
Particularly, dynamical correlation of nickel 3d bands between two different sublattices 
and that of nickel 3d and oxygen 2p bands 
are not included since DMFT neglects all the inter-atomic dynamical correlation.

To improve those problems in antiferromagnetic NiO, 
one may adopt cluster DMFT~\cite{re:rev-cluster-DMFT}, 
which is based on mapping of many electron systems in bulk 
onto single cluster impurity problem 
with including the intra-cluster dynamical Coulomb interaction 
and neglecting the inter-cluster Coulomb interaction.  
In LDA+DMFT-IPT(1), 
the extension of LDA+DMFT to LDA+cluster DMFT is consistent with 
the extension of LFT of single nickel isolated ion 
to the cluster model CI-calculation of NiO$_6$ cluster 
similar to Ref.~\onlinecite{re:NiO-fujimori2} and \onlinecite{re:NiO-vanElp}. 
The use of the cluster model CI-calculation will reproduce better positioning of occupied satellite peak 
and direct assignment of occupied main peak with $\dd^8\lbar$ final state.

From above discussion, we can conclude that 
LDA+DMFT-IPT(1) reproduces reasonable results for antiferromagnetic NiO within the DMFT scheme.
and that 
LDA+DMFT-IPT(1) is applicable to realistic compound cases and insulating cases 
as well as LDA+DMFT with nonperturbative impurity solver. 

\section{Conclusion}\label{sec:CONC}

In summary, 
we have proposed LDA+DMFT-IPT(1) 
where we use the full LDA Hamiltonian
and IPT as an impurity solver. 
We then applied LDA+DMFT-IPT(1) to ferromagnetic bcc-Fe, fcc-Ni and antiferromagnetic NiO.

For bcc-Fe and fcc-Ni case, on-site dynamical correlation effect causes 
the narrowing of the width of occupied 3d bands. 
The multiplet scattering effect of $\dd^8 \ry \dd^7$ assigned by the LFT spectrum 
causes the spin dependent $6~\eV$ satellite for fcc-Ni. 
The energy spectra and magnetic moment in LDA+DMFT-IPT(1) show a good agreement with 
previous LDA+DMFT calculations~\cite{re:Fe-Ni-UJ-paper, re:Fe-Ni-KKR+DMFT}.

For antiferromagnetic NiO case, 
on-site Coulomb interaction of nickel 3d bands enhances the hybridization between Ni-3d and O-2p bands 
and this causes the band gap of $4.3~\eV$, the charge-transfer type insulator, 
and the top of valence bands with mainly oxygen 2p bands. 
Those effects are all in good agreement with experiments. 
The drastic change of hybridization between nickel 3d and oxygen 2p bands is 
due to the use of full-LDA Hamiltonian and on-site dynamical electron correlation of nickel site within DMFT. 
The successful application of LDA+DMFT-IPT(1) to antiferromagnetic NiO implies that 
LDA+DMFT-IPT(1) is fairly applicable to larger compound cases 
with more complicated hybridization between more than two atoms.

In addition, the assignment of characteristic peaks of NiO in LDA+DMFT spectrum 
is consistent with that in the cluster-model CI calculation.~\cite{re:NiO-fujimori2} 
Thus, the assignment of the origin of peaks by using LFT spectra is a great advantage of the IPT method
 in understanding the existence of strong scattering channels in LDA+DMFT spectra.

Several remained problems in antiferromagnetic NiO, 
positions of the occupied satellite and the unoccupied main peak, 
would be solved by including the inter-atomic dynamical correlation. 

Thus, we can conclude that 
LDA+DMFT-IPT(1) is applicable to various realistic materials, 
such as both metallic and insulating cases, multi-atom (compound) cases, 
spin-polarized cases 
and strongly hybridized cases between s, p and d-bands.   


\end{document}